\documentclass[useAMS,usegraphicx,usenatbib]{mn2e}
 
\title[Abundance gradients]{Abundance gradients in spiral disks: is the gradient inversion at high redshift real?}

\author[A. Mott et al.]{A.~Mott,$^1$\thanks{E-mail: mott@oats.inaf.it}
  E.~Spitoni,$^1$ F.~Matteucci$^{1, 2}$ \\
  $^1$ Dipartimento di Fisica, Sezione di Astronomia, Universit\`a di Trieste, via G.B. Tiepolo 11, I-34131, Trieste, Italy \\
  $^2$ I.N.A.F. Osservatorio
  Astronomico di Trieste, via G.B. Tiepolo 11, I-34131, Trieste,
  Italy}

\begin{document}
\date{Accepted . ; in original form xxxx}

\pagerange{\pageref{firstpage}--\pageref{lastpage}} \pubyear{xxxx}

\maketitle

\label{firstpage}

\begin{abstract}
We compute the abundance gradients along the disk of the Milky Way by
means of the two-infall model: in particular, the gradients of oxygen
and iron and their temporal evolution. First, we explore the effects
of several physical processes which influence the formation and
evolution of abundance gradients. They are: i) the inside-out
formation of the disk, ii) a threshold in the gas density for star
formation, iii) a variable star formation efficiency along the disk,
iv) radial flows and their speed, and v) different total surface mass
density (gas plus stars) distributions for the halo. We are able to
reproduce at best the present day gradients of oxygen and iron if we
assume an inside-out formation, no threshold gas density, a constant
efficiency of star formation along the disk and radial gas flows. It
is particularly important the choice of the velocity pattern for
radial flows and the combination of this velocity pattern with the
surface mass density distribution in the halo.  Having selected the
best model, we then explore the evolution of abundance gradients in
time and find that the gradients in general steepen in time and that
at redshift $z \sim3$ there is a gradient inversion in the inner
regions of the disk, in the sense that at early epochs the oxygen
abundance decreases toward the Galactic center. This effect, which has
been observed, is naturally produced by our models if an inside-out
formation of the disk and and a constant star formation efficiency are
assumed. The inversion is due to the fact that in the inside-out
formation a strong infall of primordial gas,  contrasting chemical
  enrichment, is present in the innermost disk  regions at early
times. The gradient inversion remains also in the presence of
  radial flows, either with constant or variable speed in time, and this
  is a new result.
\end{abstract}


\section{Introduction}

The problem of understanding the formation and the evolution of the
Milky Way is fundamental to improve the knowledge of the formation of
spiral galaxies in general.  An important constraint is represented by
the abundance gradients along the thin disk of the Milky
Way. 

Abundance gradients are a feature commonly observed in many
spiral galaxies and show that the abundances of metals are decreasing
outward from the galactic centers.  A good agreement between model
predictions and observed properties of the Galaxy is generally
obtained by models that assume that the disk was formed by the infall
of gas (e.g Chiosi, 1980; Matteucci \& Fran\c cois (1989); Chiappini
et al. 1997, 2001; Boissier \& Prantzos, 1999, 2000; Fran\c cois et
al. 2004; Cescutti \& al. 2006,2007; Colavitti \& al. 2009;
Marcon-Uchida \& al. 2010 among others).  

In most of those papers the
formation timescale of the thin disk was assumed to be a function of
the galactocentric distance, leading to an inside-out scenario for the
Galaxy disk build-up. This assumption helps in reproducing the
abundance gradients along the disk.  In those papers the parameter
space of the chemical evolution models was analyzed considering
different infall and star formation rate (SFR) laws to reproduce the
observed abundance gradients.  Cescutti et al. (2007) showed that the
results obtained using a chemical evolution model with a two-infall
law and inside-out formation are in very good agreement with the
collection of data by Andrievsky et al. (2002a,b,c, 2004) and Luck et
al. (2003) on Cepheids in the Galactocentric distance range 5-17 kpc
for many of the studied elements. Colavitti et al. (2009) found that
it is impossible to fit all the disk constraints at the same time
without assuming an inside-out formation for the Galactic disk
together with a threshold in the gas density for the SFR. In
particular, the inside-out formation is important to reproduce the
slope of the abundance gradients in the inner disk, whereas the
threshold gas density is important to reproduce present-day gradients
in the outer disk.  Recently, Pilkington \& al. (2012), by means of
pure chemical models and cosmological simulations including chemistry,
have supported the conclusion that spiral disks form
inside-out. 

However, if gas infall is important, the Galactic disk is    
not adequately described by a simple multi-zone model with separated
zones (Mayor \& Vigroux 1981). To maintain consistency, radial gas
flows have to be taken into account as a dynamical consequence of
infall.  The infalling gas has a lower angular momentum than the
circular motions in the disk, and mixing with the gas in the disk
induces a net radial inflow.  Lacey \& Fall (1985) estimated that the
gas inflow velocity is up to a few km s$^{-1}$, and at 10 kpc is v$_R=
-$ 1 kms$^{-1}$.  Goetz \& K\"oppen (1992) studied numerical and
analytical models including radial flows. They concluded that radial
flows alone cannot explain the abundance gradients but are an
efficient process to amplify the existing ones. Portinari \& Chiosi
(2000) implemented the radial flows of gas in a detailed chemical
evolution model characterized by a single infall episode. 

 Recently,
Spitoni \& Matteucci (2011), and Spitoni et al. (2013) have taken into
account inflows of gas in detailed one-infall models(treating only the
evolution of the disk independently from the halo and thick disk) for
the MW and M31, respectively. Spitoni \& Matteucci (2011) tested also
the radial flows in a two-infall model but only for the oxygen.  It
was found that the observed gradient of oxygen can be reproduced if
the gas inflow velocity increases in modulus with the galactocentric
distance, in both the one-infall and two-infall models.  In contrast
to the previous papers, where the velocity patterns of the inflow were
chosen to produce a ``best fit'' model, Bilitewski \& Sch\"onrich
(2012) presented a chemical evolution model where the flow of gas is
directly linked to physical proprieties of the Galaxy like the angular
momentum budget.  The resulting velocity patterns of the flows of gas
are time dependent and show a non linear trend, always decreasing with
decreasing galactocentric distance. At a fixed galactocentric distance
the velocity flows decrease in time.
 
The abundance gradient evolution in time has been studied in several
works in the past, and in the literature, various predictions have
been made about the time evolution of metallicity gradients in
chemical evolution models: Chiappini et al. (2001) predicted a
steepening with time whereas other authors predicted an initially
negative gradient that then flattens in time (Molla \& Diaz 2005, Prantzos
\& Boissier 2000).  The reason for this discrepancy between different
models is that chemical evolution is very sensitive to the
prescriptions of the detailed physical processes that lead to the
enrichment of inner and outer disks, and the flattening or steepening
of gradients in time depends on the interplay between infall rate and
SFR along the disk and also, as discussed above, on
the presence of a threshold in the gas density for the star
formation. Different recipes of star formation or gas accretion
mechanisms can provide different abundance gradients predictions.

>From the observational point of view there have been some studies
trying to infer the temporal evolution of gradients from planetary
nebulae (PNe) of different ages (Maciel \& Costa, 2009; Maciel \&
al. 2013) but no firm conclusion could be derived.  The first
observational papers which could clarify the issue of the temporal
evolution of gradients and consequently the formation of galactic
disks, are from Cresci et al. (2010), Jones et al. (2010), Contini et
al. (2011) and Yuan et al.  (2011). In particular, Cresci \&
al. (2010) have found an inversion of the O gradient at redshift z=3
in some Lyman-break galaxies: they showed that the O abundance
decreases going toward the galactic center, thus producing a positive
gradient. This gradient inversion was already noted by Chiappini et
al.  (2001) for oxygen, and Curir et al. (2012) have adopted this
gradient inversion to study the rotation-metallicity correlation in
the thick disk found by Spagna et al. (2010), and interpreted this
correlation as the fossil of this inverted gradient at early times.

 In this paper we focus on the study of the abundance gradients in the
 Milky Way and their evolution with cosmic time with the aim of
 understanding how different physical processes can influence the
 formation and evolution of gradients in spiral galaxies in general,
 since our Galaxy can be considered as a typical spiral galaxy.  In
 particular, we will follow the evolution in time of the gradients of
 the following elements: Fe, O, C, Ne, and S. We will start from an
 updated version of the two-infall model of Chiappini et al. (2001),
 and then we will examine the processes which mainly influence the
 formation of abundance gradients: i) a differential infall law
 leading to an inside-out formation of the disk, ii) a threshold in
 the gas density in the star formation process, iii) a variable star
 formation efficiency (SFE) and iv) radial gas flows and their
 velocity pattern.  We will  finally focus our attention on the
 evolution of the abundance gradients especially at high redshift in
 order to test whether an inversion of the gradients, as measured by
 Cresci \& al. (2010)  in high redshift disk galaxies, should be
 expected for the studied chemical elements and how it can be
 interpreted.

 No chemical evolution models, with the exception of the two
  infall model of Chiappini et al.  (1997), can predict such gradient
  inversion. The novelty here is that we test whether this inversion
  is preserved in the presence of radial flows either with speed flow constant or
  variable in time.

The paper is organized as follow: in Sect. 2 we describe the chemical
evolution model used in this work, in Sect. 3 we report the
nucleosynthesis prescriptions, in Sect. 4 we present the
implementation of the radial inflow of gas in the chemical model, and
in Sect. 5 observational data are shown.  In Sect. 6 we report and
discuss our model results concerning the present day gradients and the
ones at high redshift.  Finally we draw the main conclusions in
Sect. 7.

\section{The chemical evolution model for the Milky Way}

The two-infall model considered here is an updated version of the
Chiappini et al. (1997; 2001) model. The Galaxy is assumed to have formed by
means of two main infall episodes: the first formed the halo and the
thick disk, the second the thin disk.  The time-scale for the
formation of the halo-thick disk is 0.8 Gyr and the entire formation
period does not last more than 2 Gyr. The time-scale for the thin disk
is much longer, 7 Gyr in the solar vicinity, implying that the
infalling gas forming the thin disk comes mainly from the
intergalactic medium and not only from the halo.  Moreover, the
formation timescale of the thin disk is assumed to be a function of
the galactocentric distance, leading to an inside out scenario for the
Galaxy disk build-up (Matteucci \& Fran\c cois 1989).  The galactic
thin disk is approximated by several independent rings, 2 kpc wide,
without exchange of matter between them.

 The main characteristic of the two-infall model is an almost
 independent evolution between the halo and the thin disk (see also
 Pagel \& Tautvaisiene 1995).  A threshold gas density of
 $7M_{\odot}pc^{-2}$ in the star formation process (Kennicutt 1989,
 1998, Martin \& Kennicutt 2001) is also adopted for the disk.

The equation below describes the time evolution of $G_{i}$, namely the
mass fraction of the element $i$ in the gas:

\begin{displaymath}
\dot{G_{i}}(r,t)=-\psi(r,t)X_{i}(r,t)
\end{displaymath}
\smallskip
\begin{displaymath}
+\int\limits^{M_{Bm}}_{M_{L}}\psi(r,t-\tau_{m})Q_{mi}(t-\tau_{m})\phi(m)dm
\end{displaymath}
\begin{displaymath}
+A_{Ia}\int\limits^{M_{BM}}_{M_{Bm}}\phi(M_{B})\left[\int\limits_{\mu_{m}}^{0.5}f(\mu)\psi(r,t-\tau_{m2})Q^{SNIa}_{mi}(t-\tau_{m2})d\mu\right]dM_{B}
\end{displaymath}
\begin{displaymath}
+(1-A_{Ia})\int\limits^{M_{BM}}_{M_{Bm}}\psi(r,t-\tau_{m})Q_{mi}(t-\tau_{m})\phi(m)dm
\end{displaymath}
\begin{displaymath}
+\int\limits^{M_{U}}_{M_{BM}}\psi(r,t-\tau_{m})Q_{mi}(t-\tau_{m})\phi(m)dm
\end{displaymath}
\smallskip
\begin{equation}
+X_{A_{i}}A(r,t)
\label{evol}
\end{equation}
where $X_{i}(r,t)$ is the abundance by mass of the element $i$ and
$Q_{mi}$ indicates the fraction of mass restored by a star of mass $m$
in form of the element $i$, the so-called ``production matrix'' as
originally defined by Talbot \& Arnett (1973). We indicate with
$M_{L}$ the lightest mass which contributes to the chemical enrichment
and it is set at $0.8M_{\odot}$; the upper mass limit, $M_{U}$, is set
at $100M_{\odot}$.

The SFR $\psi(r,t)$ is defined as:
\begin{equation}
\psi(r,t)=\nu\left(\frac{\sigma(r,t)}{\sigma(r_{\odot},t)}\right)^{2(k-1)}
\left(\frac{\sigma(r,t_{Gal})}{\sigma(r,t)}\right)^{k-1}G^{k}(r,t),
\end{equation}
 where $\nu$ is the efficiency of the star formation process and is
 set to be 2 Gyr$^{-1}$ for the Galactic halo and 1 Gyr$^{-1}$ for the
 disk.  $\sigma(r,t)$ is the total surface mass density,
 $\sigma(r_{\odot},t)$ the total surface mass density at the solar
 position, $G(r,t)$ the surface gas density.  The age of the Galaxy is
 $t_{Gal}=14$ Gyr and $r_{\odot}=8$ kpc is the solar galactocentric
 distance (Reid 1993). The gas surface exponent, $k$, is set equal to
 1.5 (Kennicutt 1989).  With these values of the parameters, the
 observational constraints, in particular in the solar vicinity, are
 well fitted.  Below a critical threshold of the gas surface density
 ($7M_{\odot}pc^{-2}$), we assume no star formation.  This naturally
 produces a hiatus in the SFR between the halo-thick disk and the thin
 disk phase, as extensively discussed in Chiappini et al. (2001).

For the IMF, we use that of  Scalo (1986), constant in time and space.
$\tau_{m}$ is the evolutionary lifetime of stars as a function of their mass {\it m} (Maeder \& Maynet 1989).

The Type Ia SN rate has been computed following Greggio \& Renzini (1983)
and  Matteucci \& Greggio (1986) and it is expressed as:
\begin{equation}
R_{SNeIa}=A_{Ia}\int\limits^{M_{BM}}_{M_{Bm}}\phi(M_{B})\left[ \int\limits^{0.5}_{\mu_{m}}f(\mu)\psi(t-\tau_{M_{2}})d\mu \right]
dM_{B}
\end{equation}
where $M_{2}$ is the mass of the secondary, $M_{B}$ is the total mass of the binary
system, $\mu=M_{2}/M_{B}$, $\mu_{m}=max\left[M_{2}(t)/M_{B},(M_{B}-0.5M_{BM})/M_{B}\right]$, 
$M_{Bm}= 3 M_{\odot}$, $M_{BM}= 16 M_{\odot}$. The IMF is represented by $\phi(M_{B})$
and refers to the total mass of the binary system for the computation of the Type Ia SN rate,
$f(\mu)$ is the distribution function for the mass fraction of the secondary:
\begin{equation}
f(\mu)=2^{1+\gamma}(1+\gamma)\mu^{\gamma}  
\end{equation}
with $\gamma=2$; $A_{Ia}$ is the fraction of systems in the
appropriate mass range, which can give rise to Type Ia SN events. This
quantity is fixed to 0.05 by reproducing the observed Type Ia SN rate
at the present time (Mannucci et al. 2005). Note that in the case of
the Type Ia SNe the``production matrix'' is indicated with
$Q^{SNIa}_{mi}$ because of its different nucleosynthesis contribution
(for details see Matteucci \& Greggio 1986 and Matteucci 2001).

The term $A(r,t)$ represents the accretion term and is defined as:
\begin{equation}
A(r,t)= a(r) e^{-t/ \tau_{H}(r)}+ b(r) e^{-(t-t_{max})/ \tau_{D}(r)}.
\end{equation}
The quantities $X_{A_{i}}$ are the abundances in the infalling
material, which is assumed to be primordial, while $t_{max}=1$ Gyr is
the time for the maximum infall on the thin disk, $\tau_{H}= 0.8$ Gyr
is the time scale for the formation of the halo thick-disk and
$\tau_{D} (r)$ is the timescale for the formation of the thin disk and
is a function of the galactocentric distance (inside-out formation,
Matteucci and Fran\c cois, 1989; Chiappini et al. 2001).
 
In particular, we assume that:
\begin{equation}
\tau_{D}=1.033 r (\mbox{kpc}) - 1.267 \,\, \mbox{Gyr}.
\end{equation}
Finally, the coefficients $a(r)$ and $b(r)$ are obtained  by imposing a fit  
 to the observed current total surface mass density in the thin disk
as a function of galactocentric 
distance given by:
\begin{equation}
\sigma(r)=\sigma_{0}e^{-r/r_{D}},
\end{equation}
where $\sigma_{0}$=531 $M_{\odot}$ pc$^{-2}$ is the central total
surface mass density and $r_{D}= 3.5$ kpc is the scale length.

\section{Nucleosynthesis prescriptions}

For the nucleosynthesis prescriptions of the Fe and the other elements
(namely O, S, Si, Ca, Mg, Sc, Ti, V, Cr, Zn, Cu, Ni, Co and Mn ), we
adopted those suggested in Francois et al. (2004).  They compared
theoretical predictions for the [el/Fe] vs. [Fe/H] trends in the solar
neighborhood for the above mentioned elements and they selected the
sets of yields required to best fit the data but not for iron.  In
particular, for the yields of SNe II they found that the Woosley \&
Weaver (1995) ones provide the best fit to the data.  No modifications
are required for the yields of Ca, Fe, Zn and Ni as computed for solar
chemical composition. For oxygen, the best results are given by the
Woosley \& Weaver (1995) yields computed as functions of the
metallicity.  For the other elements, variations in the predicted
yields are required to best fit the data (see Francois et al. (2004)
for details).

Concerning the yields from
Type SNeIa, revisions in the theoretical yields by Iwamoto et
al.(1999) are required for Mg, Ti, Sc, K, Co, Ni and Zn to best fit
the data.  The prescription for single low-intermediate mass stars is
by van den Hoek \& Groenewegen (1997), for the case of the mass loss
parameter, which varies with metallicity (see Chiappini et al. 2003,
model 5). Here we will concentrate on the evolution of O and Fe for which no
corrections are needed.

\section{The implementation of the radial inflow}

  We implement radial inflows of gas in our reference model
following the prescriptions described in Spitoni \& Matteucci (2011).

We define the $k$-th shell in terms of the galactocentric radius $r_k$,
its inner and outer edge being labeled as $r_{k-\frac{1}{2}}$ and
$r_{k+\frac{1}{2}}$.  Through these edges, gas inflow occurs with velocity
v$_{k-\frac{1}{2}}$ and v$_{k+\frac{1}{2}}$, respectively. The flow
velocities are assumed to be positive outward and  negative inward.

 Radial inflows  with a flux $F(r)$, contribute to altering 
the gas surface density $\sigma_{gk}$  in the $k$-th shell in according to
\begin{equation}
\label{dsigmarf1}
\left[ \frac{d \sigma_{g k}}{d t} \right]_{rf} = 
   - \frac{1}{\pi \left( r^2_{k+\frac{1}{2}} - r^2_{k-\frac{1}{2}} \right) }
   \left[ F(r_{k+\frac{1}{2}}) - F(r_{k-\frac{1}{2}}) \right],
\end{equation}
where the gas flow at $r_{k+\frac{1}{2}}$ can be written as
\begin{small}
\begin{equation}
\label{flux3}
F(r_{k+\frac{1}{2}}) = 2 \pi r_{k+\frac{1}{2}} \, v_{k+\frac{1}{2}} \left[\sigma_{g (k+1)} \right].
\end{equation}
\end{small}

We take the inner edge of the
$k$-shell, $r_{k-\frac{1}{2}}$, at the midpoint between the
characteristic radii of the shells $k$ and $k-1$, and similarly for
the outer edge $r_{k+\frac{1}{2}}$

$r_{k-\frac{1}{2}}= (r_{k-1}+ r_k)/2$, and $r_{k+\frac{1}{2}}= (r_{k}+
r_{k+1})/2$. We find that
\begin{equation}
\left(r^{2}_{k+\frac{1}{2}} -r^{2}_{k-\frac{1}{2}} \right)=\frac{ r_{k+1}-r_{k-1}}{2}\left( r_k+\frac{r_{k-1}+r_{k+1}}{2}\right).  
\label{rs}
\end{equation}

Inserting eqs. (\ref{flux3}) and (\ref{rs}) into
eq. (\ref{dsigmarf1}), we obtain the radial flow term to be added into
eq. (\ref{evol})

\begin{equation}
\left[ \frac{d}{dt} G_i(r_k,t) \right]_{rf} = -\, \beta_k \,
G_i(r_k,t) + \gamma_k \, G_i(r_{k+1},t),
\end{equation}
where $\beta_k$ and $\gamma_k  $ are, respectively:

\begin{equation}
\beta_k =  - \, \frac{2}{r_k + \frac {r_{k-1} + r_{k+1}}{2}}  
	\times \left[ v_{k-\frac{1}{2}}
	    \frac{r_{k-1}+r_k}{r_{k+1}-r_{k-1}}  \right]  
\end{equation}

\begin{equation}
\gamma_k =  - \frac{2}{r_k + \frac {r_{k-1} + r_{k+1}}{2}} 
	     \left[ v_{k+\frac{1}{2}}
	     \frac{r_k+r_{k+1}}{r_{k+1}-r_{k-1}} \right] 
	     \frac{\sigma_{(k+1)}}{\sigma_{k}},
\end{equation}
where $\sigma_{(k+1)}$ and $\sigma_{k}$ are the present time total
surface mass density profile at the radius $r_{k+1}$ and $r_{k}$,
respectively.  We assume that there are no flows from the outer parts
of the disk where there is no SF. In our implementation of the radial
inflow of gas, only the gas that resides inside the Galactic disk
within the radius of 20 kpc can move inward by radial inflow.

\section{Observational Data}

The data set we used are those presented by Luck \& Lambert
(2011) which contain a spectroscopic investigation of 238 Cepheids in
the northern sky in the Galactocentric distance range 5-17 kpc. Among
these stars, about 150 are new to the study of the galactic abundance
gradients, while the others are taken from the previous work by Luck
et al. (2011), although these latter stars have been re-observed in
order to have a final, complete and homogeneous set of data.  The
solar abundances adopted in Luck \& Lambert (2011) are the same
adopted in this paper: for the oxygen it is $log (\epsilon_O) = 8.69 \pm
0.05$ (Asplund et al.  2005), whereas for the iron it is $log
(\epsilon_{Fe}) = 7.50 \pm 0.04$ (Asplund et al. 2009).

\begin{figure} 
	    \includegraphics[scale=0.4]{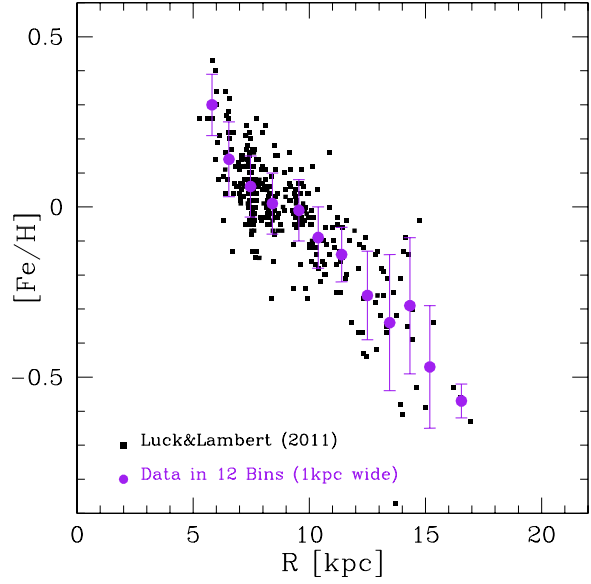} 
    \caption{Radial abundance gradient for iron from observations of
      Cepheids. The data (black filled squares) are taken by Luck \&
      Lambert (2011). The purple dots are the bins in which the data
      are divided, with their uncertainties. Each bin represents the
      mean abundance value in the stars located in proximity of each
      kpc, whereas the error bars are the standard deviation computed
      in each bin.}
		\label{ferro_dat}
\end{figure}

\begin{figure}
  \includegraphics[scale=0.4]{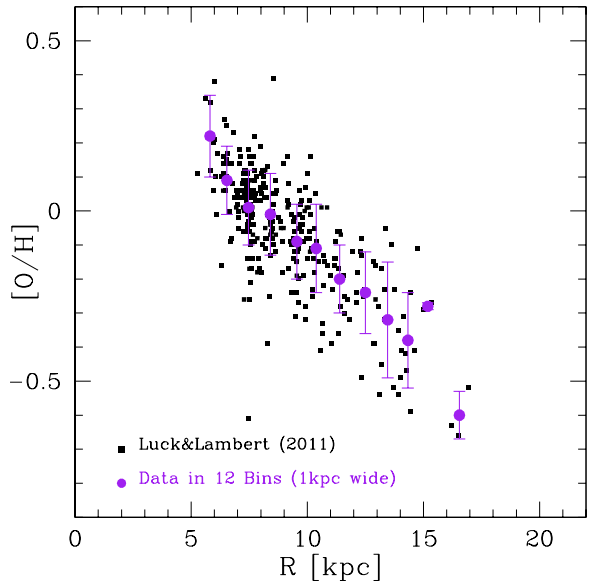} 
    \caption{Radial abundance gradient for oxygen from observations of
      Cepheids. The data (black filled squares) are taken by Luck \&
      Lambert (2011). The purple dots are the bins in which the data
      are divided, with their uncertainties. Each bin represents the
      mean abundance value in the stars located in proximity of each
      kpc, whereas the error bars are the standard deviation computed
      in each bin.}
		\label{oss_dat}
\end{figure}

To better understand the trend of the data, we divided them in twelve
bins as functions of Galactocentric distance, each 1 kpc wide. In this
way it is easy to compare the observational data with the predictions
of our chemical evolution models. In each bin we computed the mean
abundance value and the standard deviation for the studied element. In
Figs. \ref{ferro_dat} and \ref{oss_dat} we report the whole collection
of data together with the adopted bin division, for the iron and the
oxygen, respectively.

\begin{figure} 
    \includegraphics[scale=0.4]{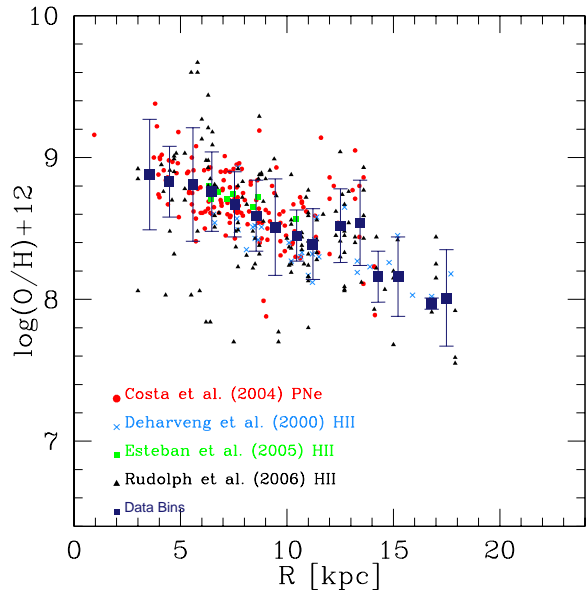} 
    
    \caption{Radial abundance gradient for oxygen from
      observations. The data are taken by Deharveng et al. (2000)
      (light-blue crosses), Rudolph et al. (2006) (black triangles),
      Esteban et al. (2005) (green squares) for the HII regions and by
      Costa et al. (2004) (red filled circles) for the PNe.}
		\label{PNe_HII}
\end{figure}

We also compare our model results with the data from HII regions and
PNe.  In this paper we adopt the HII region data by Deharveng et
al. (2000), Esteban et al.  (2005) and Rudolph et al. (2006), who
analyzed the Galactic HII regions, and Costa et al. (2004) data
for planetary nebulae (PNe).  As for the Cepheids, the data for the
sample PNe+HII regions have been treated in the same way, dividing
them into twelve bins, each 1 kpc wide, as shown in Fig. \ref{PNe_HII}.

We have computed the best fits to the abundance gradients for O and Fe
from Cepheids on one hand,  and from HII regions and PNe on the other hand. 
In particular, by using
the least square method, the abundance gradients from the data of
Cepheids of Luck \& Lambert (2011) are:

\begin{equation}
\frac{d\left[\mbox {Fe/H}\right]}{dr}=-0.063\pm  0.003\ \mbox {dex/kpc\ \ (5-17\ kpc)},
\end{equation}
\begin{equation}
\frac{d\left[\mbox {O/H}\right]}{dr}\,\,=-0.057\pm 0.003\ \mbox{dex/kpc\ \  (5-17\ kpc)}. 
\end{equation}  
 
The data set for  HII regions and PNe suggest for the oxygen the follow gradient: 

\begin{equation}
\frac{d\left(log\left(\frac{O}{H}\right)+12\right)}{dr}=-0.059\pm  0.005\ \mbox {dex/kpc\ \ (2-18\ kpc)}.
\end{equation}

It is worth noting that the Cepheid gradient obtained by Luck \&
Lambert (2011) is in excellent agreement with the gradient derived for
HII regions and PNe.

  In this work we also discuss the formation and the temporal
  evolution of the abundance gradients of different elements. To study
  their evolution in time, we need to compare our model results to the
  abundances as functions of the Galactocentric distance at various
  evolutionary stages. To compare our predictions we then need to find
  abundance data samples at redshift higher than zero. Unfortunately,
  abundance gradients at earlier times than the present one are very
  uncertain, such as for example those derived from old PNe. Some PNe
  could be in fact as old as 8-10 Gyr (PNe of type III) , whereas
  others are relatively young (PNe of type I). Maciel \& Costa (2009)
  and Maciel \& al. (2013) suggested a flattening of the gradient of O
  in the last 6-8 Gyr but the error on the estimated ages are probably
  too large to derive any firm conclusion. Recent data from Cresci et
  al. (2010) report oxygen abundances across three star-forming
  galaxies at redshift z $\sim$ 3.  In this study, we consider two
  Lyman-Break galaxies, $SSA22a - C16$ at $z = 3.065$ and $SSA22a - M
  38$ at $z = 3.288$ which have similar properties to our Milky Way:
  each of them are star forming galaxies with masses, found assuming
  an exponential disk mass model to reproduce the dynamical properties
  of these sources, in the range $10^{10}-2\cdot10^{12}\ M_{\odot}$,
  comparable with the mass of our Galaxy (see Table 1). In particular,
  in Table 1 we show the name of the galaxies, their redshifts, the
  estimated masses, the O abundances integrated over the whole
  galaxies as well as the O abundances in the low and high metallicity
  regions.  The most striking result of this study is the existence of
  a large positive gradient in the oxygen abundance, in other words
  the O abundance in these galaxies seem to decrease towards the
  galactic centre. Although the data refer to regions in small
  galactocentric distance ranges (see Table 2) they clearly indicate a
  ``gradient inversion'' at redshift z=3.  In particular, in Table 2
  are reported the abundance gradients for the two Lyman Break
  galaxies obtained by considering the location (in kpc from the
  galactic centers) of the low and high metallicity regions, as
  defined in Cresci et al. (2010).  In this paper we aim at
  understanding if this inversion is predicted by chemical evolution
  models and what can be its physical meaning.

\begin{table}
\centering  
\renewcommand\arraystretch{1.35}	
\resizebox{0.48\textwidth}{!}{														
\begin{tabular}{l|c|c|c|c|c} 									
\hline 																		
\multicolumn{1}{c|}{\textbf{Source}}&\multicolumn{1}{c|}{\textbf{Redshift}}&\multicolumn{1}{c|}{\textbf{Mass}}&\multicolumn{3}{c}{\textbf{$log(O/H)+12$}}\\
\multicolumn{1}{c|}{\textbf{Name}}&\multicolumn{1}{c|}{$z$}&\multicolumn{1}{c|}{\textbf{$log\left(\frac{M_{\star}}{M_{\odot}}\right)$}}&\multicolumn{1}{c}{(tot)}&\multicolumn{1}{c}{(low Z)}&\multicolumn{1}{c}{(high Z)}\\
\hline
\multicolumn{1}{l|}{\textbf{SS22a-C16}}&\multicolumn{1}{c|}{3.065}&\multicolumn{1}{c|}{$10.68^{+0.16}_{-0.54}$}&\multicolumn{1}{c}{$8.36^{+0.06}_{-0.06}$}&\multicolumn{1}{c}{$8.18^{+0.13}_{-0.14}$}&\multicolumn{1}{c}{$8.52^{+0.14}_{-0.07}$}\\
\multicolumn{1}{l|}{\textbf{SS22a-M38}}&\multicolumn{1}{c|}{3.288}&\multicolumn{1}{c|}{$10.86^{+0.18}_{-0.41}$}&\multicolumn{1}{c}{$8.26^{+0.09}_{-0.11}$}&\multicolumn{1}{c}{$7.84^{+0.22}_{-0.23}$}&\multicolumn{1}{c}{$8.59^{+0.05}_{-0.07}$}\\
\hline
\end{tabular}}
\caption{Oxygen abundances by Cresci et al. (2010) at z=3. $z$ is the source redshift, $M_{\star}$ the stellar mass, while the
last three columns show the metallicity integrated for the whole galaxy, for
the low metallicity region and for the high metallicity regions respectively,
in units of $12+log(O/H)$. The uncertainties are $1\ \sigma$ confidence interval.} 			
\label{cresci}
\end{table}

\begin{table}
\centering  
\renewcommand\arraystretch{1.35}	
\resizebox{0.48\textwidth}{!}{														
\begin{tabular}{l|c|c|c} 			 						
\hline 																		
\multicolumn{1}{c|}{\textbf{Name}}&\multicolumn{1}{|c|}{\textbf{z}}&\multicolumn{1}{|c|}{$log\left(\frac{M_{\star}}{M_{\odot}}\right)$}&\multicolumn{1}{|c}{$\frac{d[log(O/H)+12]}{dRg}$ (range [kpc])}\\
\hline
\multicolumn{1}{l|}{\textbf{SS22a-C16}}&\multicolumn{1}{|c|}{\textbf{3.065}}&\multicolumn{1}{|c|}{$10.68^{+0.16}_{-0.54}$}&\multicolumn{1}{|c}{+0.13 \footnotesize(1.9-4.6)}\\
\multicolumn{1}{l}{\textbf{SS22a-M38}}&\multicolumn{1}{|c|}{\textbf{3.288}}&\multicolumn{1}{|c|}{$10.86^{+0.18}_{-0.41}$}&\multicolumn{1}{|c}{+0.50 \footnotesize(1.9-3.4)}\\
\hline
\end{tabular}}
\caption{Oxygen abundance gradients from two different metallicity zones in each galaxy at $z\sim3$.} 
\label{gradienti_cresci}
\end{table}

\section{Model results}

 We use as a reference model an updated version of the Chiappini et
 al. (2001) model which corresponds also to the best model in Cescutti et al. (2007). First of all, we investigate in details the effects
 of several physical processes on abundance gradients:
\begin{itemize}
\item The inside-out formation of the thin disk, obtained by 
varying $\tau_D$ as a
function of the galactocentric distance;
\item The presence of a threshold in the gas density for star formation;
\item A variable efficiency of star formation with the galactocentric distance;
\item The radial gas flows and their velocity pattern
\end{itemize}

\begin{table*}

\caption{The list of the models described in this work.}
\scriptsize

\label{models}
\begin{center}
\begin{tabular}{c|cccccc}
  \hline
\hline
\\
 Model &Threshold [M$_{\odot}$ pc$^{-2}$]& $\tau_d$ (I-O) [Gyr] & $\sigma_{h}(r)$ [M$_{\odot}$ pc$^{-2}$] &SFE $\nu$ [Gyr$^{-1}$]&Radial inflow  \\
\hline

S-A& 7 (Thin Disk) &  1.033 R (kpc) -1.27 Gyr &17&1&/\\
& 4 (Halo-Thick Disk)&&&\\
\hline
S-B& 7 (Thin Disk) &  1.033 R (kpc) -1.27 Gyr &17&$\nu(R) \propto R^{-1}$&/\\
& 4 (Halo-Thick Disk)&&& Colavitti et al. (2009)&\\
\hline
S-C& / &  1.033 R (kpc) -1.27 Gyr &17&1&/\\
\hline

S-E& 7 (Thin Disk) &  3 Gyr &17&1&/\\
& 4 (Halo-Thick Disk)&&&\\
\hline

S-G& / &  1.033 R (kpc) -1.27 Gyr &17 if R$\leq 8$ kpc &1&/\\
& && 0.01 if R$\geq 10$  &&\\
\hline

R-A& 7 (Thin Disk) &  1.033 R (kpc) -1.27 Gyr &17&1&velocity pattern I\\
& 4 (Halo-Thick Disk)&&&&\\
\hline
R-B& 7 (Thin Disk) &  1.033 R (kpc) -1.27 Gyr &17&$\nu(R) \propto R^{-1}$&velocity pattern I\\
& 4 (Halo-Thick Disk)&&&  Colavitti et al. (2009)&\\

 \hline
R-C& 7 (Thin Disk) &  1.033 R (kpc) -1.27 Gyr &17&1 &velocitiy pattern II\\
& 4 (Halo-Thick Disk)&& &&\\

 \hline
R-D& / &  1.033 R (kpc) -1.27 Gyr &17 if R$\leq 8$ kpc&1 &velocitiy pattern III\\
&  &&      0.01 if R$\geq 10$          &&\\

\hline
\end{tabular}
\end{center}

\end{table*}

In Table 3 all the models considered in this work are reported.  The
model names are shown in the first column: the letter ``S'' means that
it is a static model, namely without any radial inflow in the disk,
while ``R'' stands for the models that take into account the radial
flows. If a threshold is present in the model, its values for the halo
and thin disk phases are shown in the second column, expressed in
$M_{\odot} pc^{-2}$. The inside-out scenario is expressed by a linear
variation of the timescale of infall $\tau_D(R)$, as shown in eq. (6),
and if this assumption is absent, the timescale is set to be constant
having a value of 3 Gyr.  In column 4 is indicated the assumed total
(stars plus gas) surface mass density profile of the halo in
$M_{\odot} pc^{-2}$ (see Chiappini et al. 2001).  In column 5 the SFR
is indicated.  The variable SFE $\nu(R)$ assumes
different values at different galactocentric distances, as in the work
by Colavitti et al. (2009). This approach is not new since a variable
$\nu$ has already been proposed on the basis of large-scale
instabilities in rotating disks (e.g. Boissier \& Prantzos 1999;
Boissier et al. 2001).  We assume a star formation efficiency (SFE)
that increases towards the innermost regions of the disk.  For $R \leq
14$ kpc the efficiency is $\nu \propto R^{-1}$, while in the outer
parts of the disk we assume a constant value of $\nu = 0.03$
Gyr$^{-1}$; the reason is that in these outer regions the gas density
is very small and decreasing exponentially, so assuming a decreasing
$\nu$ would lead to a very small value for this parameter and
consequently to negligible star formation. 

 Lacey and Fall (1985)
studied the Galactic chemical evolution in the presence of radial
inflow of gas and demonstrated that such flow enhances the metallicity
gradient within the disk. As shown in Sch\"onrich and Binney (2009)
and Spitoni \& Matteucci (2011), to reproduce the data it is necessary
to adopt a variable velocity for the radial gas flow.  In this
situation, each ring has its own velocity inflow.  Here we use several
velocity patterns, shown in Fig. \ref{patternmott}. The velocities in
these patterns always decrease inwards (in agreement with Edmunds \&
Greenhow 1995), and in modulus they span the range 0-5 km sec$^{-1}$ ,
in accordance with the results of Sch\"onrich and Binney (2009) and
other previous works in literature (Wong et al. 2004; Lacey \& Fall
1985; and Tinsley 1980).

  Our radial inflow patterns  are in
  accordance with the ones computed by Bilitewski \& Sch\"onrich
  (2012), imposing the conservation of the angular momentum. Fu et
  al. (2013) in semi-analytic models of disk galaxy formation
  have taken into account  a radial inflow of gas in agreement  with this work and
  Spitoni \& Matteucci (2011). Moreover, Wang \& Zhao (2013) presented
  a chemical evolution model for the Milky Way including stellar
  migration and radial inflows of gas, and again the chosen patterns
  are in agreement with  our models.

\subsection{The present day gradient results }
In Figs. \ref{parametri14ossigeno} and \ref{parametri14ferro} we show
our results for the oxygen and iron present day gradient,
respectively, compared with the data from Cepheids.
The model labeled S-A is our reference model and it is just an updated
version of the Chiappini \& al. (2001) Model B. The Galactic disk in
this model forms inside-out; this means that the inner parts are
assembled before the outer ones.  This kind of formation is quite
successful in reproducing the main features of the Milky Way as well
as of the external galaxies, especially concerning abundance gradients
(e.g. Matteucci \& Fran\c cois, 1989; Chiappini et al. 2001; Prantzos
\& Boissier 2000; Pilkington et al. 2012).

\begin{figure}
	  \centering   
    \includegraphics[scale=0.4]{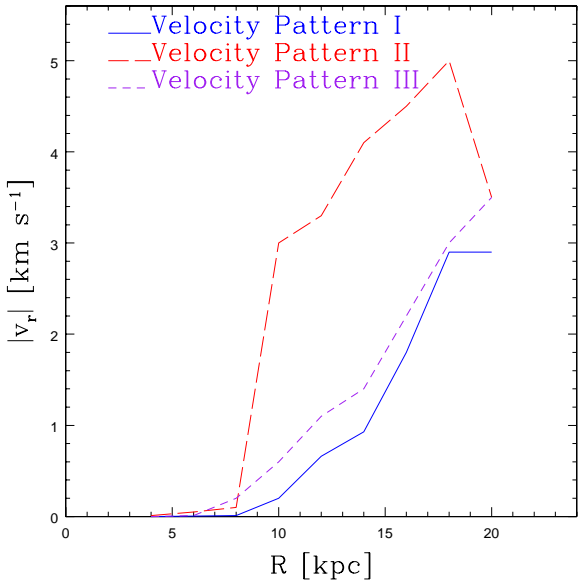} 
    \caption{Radial inflow velocity profiles adopted in this paper (see text).}
		\label{patternmott}
\end{figure}

\begin{figure}
	  \centering   
    \includegraphics[scale=0.4]{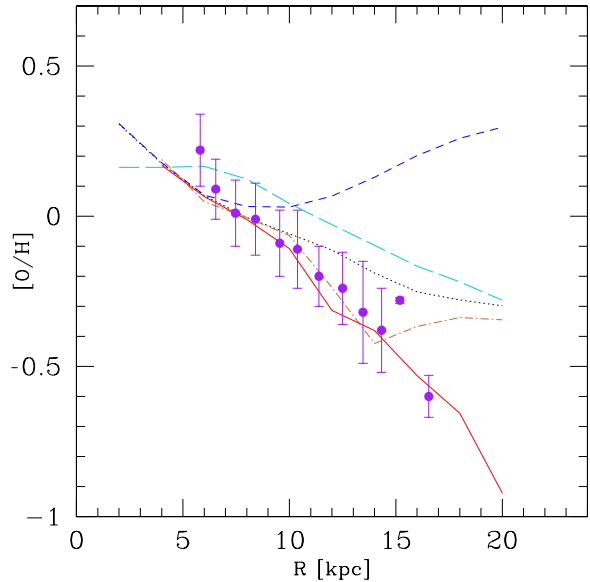} 
    \caption{Effect on the oxygen abundance gradients of the four
      parameters that characterize the chemical evolution models where
      the halo surface density is fixed at the value of $\sigma_H=17$
      M$_{\odot}$ pc$^{-2}$: threshold (model S-C, blue short dashed
      line), Inside-Out (model S-E, light-blue long dashed line),
      variable SFE (model S-B, brown dashed
      line) and radial flows (model R-A, red solid line). The black
      dotted line represents the static model by Chiappini et
      al. (1997), model S-A with Inside-Out, threshold, constant SFE
      and without inflow. The data are by Luck \& Lambert (2011).}
\label{parametri14ossigeno}
\end{figure}  

If instead we do not assume an inside-out formation for the thin disk,
and keep constant the timescale of infall $\tau_D$ , the present day
abundance gradients provided by the model are too flat in the inner
part of the disk, as shown by the model S-E in both
Figs. \ref{parametri14ossigeno} and \ref{parametri14ferro}, even if a
threshold in the gas density is assumed.  In fact, the threshold
influences mostly the outer gradient.

\begin{figure} 
	  \centering   
    \includegraphics[scale=0.4]{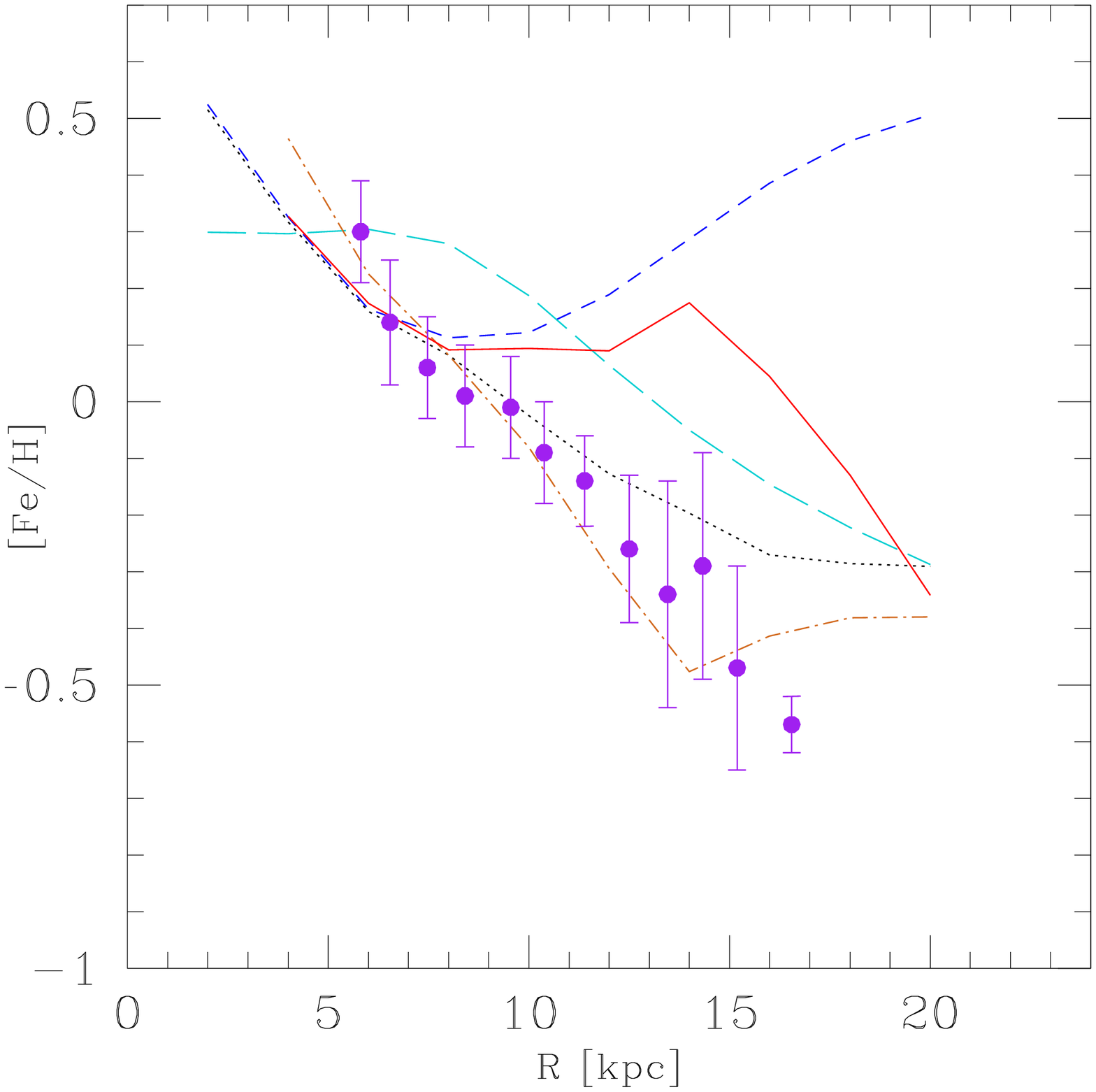} 
    \caption{Effect on the iron gradients of the four parameters that
      characterize the chemical evolution models where the halo
      surface density is fixed at the value of $\sigma_{h}=17$
      M$_{\odot}$ pc$^{-2}$: threshold (model S-C, blue short dashed
      line), Inside-Out (model S-E, light-blue long dashed line),
      variable SFE (model S-B, brown dashed
      line) and radial flows (model R-A, red solid line). The black
      dotted line represents the static model by Chiappini et
      al. (1997), model S-A with Inside-Out, threshold, constant SFE
      and without inflow. The data are by Luck \& Lambert (2011).}
		\label{parametri14ferro}
\end{figure}

\begin{figure} 
	  \centering   
    \includegraphics[scale=0.4]{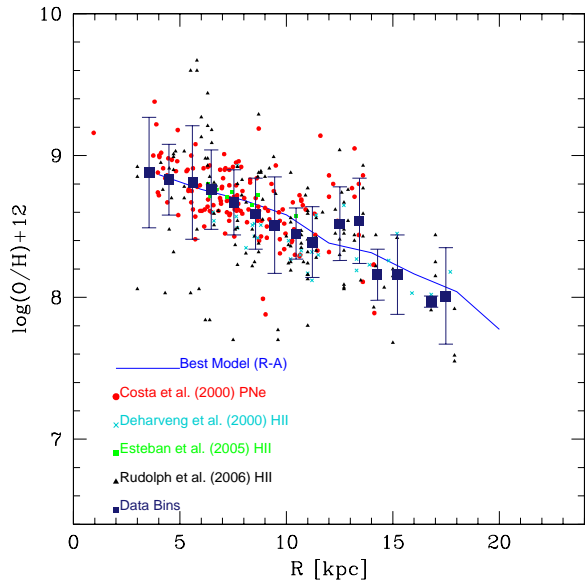} 
    \caption{ Comparison between the data set composed by HII regions
      and PNe for the oxygen, and the R-A model predictions. This
      model assumes Inside-Out formation, a threshold in the gas
      density, and radial inflow of gas with the velocity pattern
      taken by Spitoni \& Matteucci (2011)}
		\label{Pnebest}
\end{figure}

From Figs. \ref{parametri14ossigeno} and \ref{parametri14ferro} it is
clear that without a threshold in the standard static model (model
S-C), even if an inside-out formation is considered, the gradient is
too flat between 6-12 kpc and in the outer parts of the disk it even
increases, clearly at variance with the observational data. The reason
why the threshold steepens the gradient in the outermost regions is
because every time the gas density goes below the critical value (and
it is easier to happen in the outer regions), the star formation stops
and starts again only when the gas has gone over the threshold thanks
to gas infalling and/or restored by dying stars. Actually, at
distances larger than 15-16 kpc the gas density is almost always below
the assumed threshold so the abundances are very low, due to the
frequent suppression of star formation. This is in agreement with what
found in Chiappini et al. (2001).  Thus, we can conclude that a
threshold in the gas density seems to be necessary to have the right
trend of the gradients in the outer parts of the disk in a static
model with the inside-out formation.  The model which assumes a
variable SFE, a threshold, and inside-out formation (model S-B),
provides a good fit to the observed present day abundance gradients
from 4 to 14 kpc. However, beyond this distance the gradient predicted
by the models is too flat and inconsistent with the observations. This
is due to the fact that we have assumed a constant and very low ($\nu=
0.03$ Gyr$^{-1}$) SFE at distances $> 14$ kpc.  Finally, the model R-A
assumes the presence of a radial gas inflow following the velocity
pattern (pattern I) similar to that adopted in Spitoni \& Matteucci
(2011) and shown in Fig. \ref{patternmott}.  Using this velocity
pattern for the radial inflow, we can see from
Fig. \ref{parametri14ossigeno} that the oxygen gradient provided by
the model R-A (-0.058 $\pm$ 0.005) is steeper than the one predicted
by the static model S-A (-0.030 $\pm$ 0.002), and in better agreement
with the observational data. In Fig. \ref{Pnebest} we show that the
model R-A perfectly fits also the gradient observed in the HII regions
and PNe.

However, it is clear from Fig. \ref{parametri14ferro} that the Fe
gradient from model R-A does not fit the present time gradient, at
least in the outer regions, and in any case the fit is worse than in
the static model S-A. In particular, in the presence of radial flows
with the velocity pattern I the Fe gradient flattens between 9 and 14
kpc. So, with this kind of radial flows the O gradients steepens,
whereas that of Fe flattens.

 \begin{figure}
	  \centering   
    \includegraphics[scale=0.4]{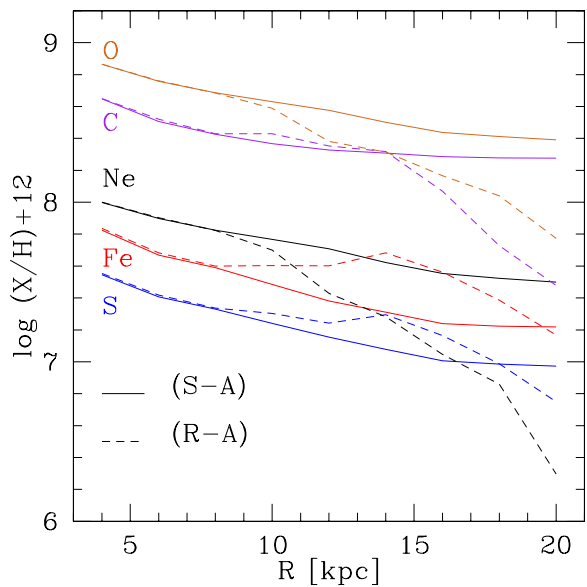} 
    \caption{Present day abundance gradients for O, C, Ne, Fe, and S
      for the static model S-A (solid lines) and for the model with
      radial inflow of gas R-A (dashed lines).}
		\label{not_l}
\end{figure}

\begin{figure} 
	  \centering   
    \includegraphics[scale=0.4]{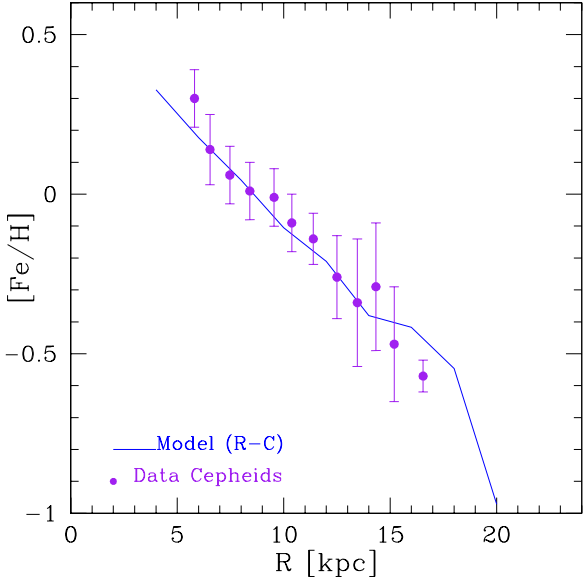} 
    \caption{Iron abundance gradient: comparison between observations
      and model (R-C) predictions. This fit is obtained using the
      velocity pattern (II) shown in Fig. \ref{patternmott}. The data
      are taken from Luck \& Lambert (2011).}
		\label{bestFe}
\end{figure}

\begin{figure}
	  \centering   
    \includegraphics[scale=0.4]{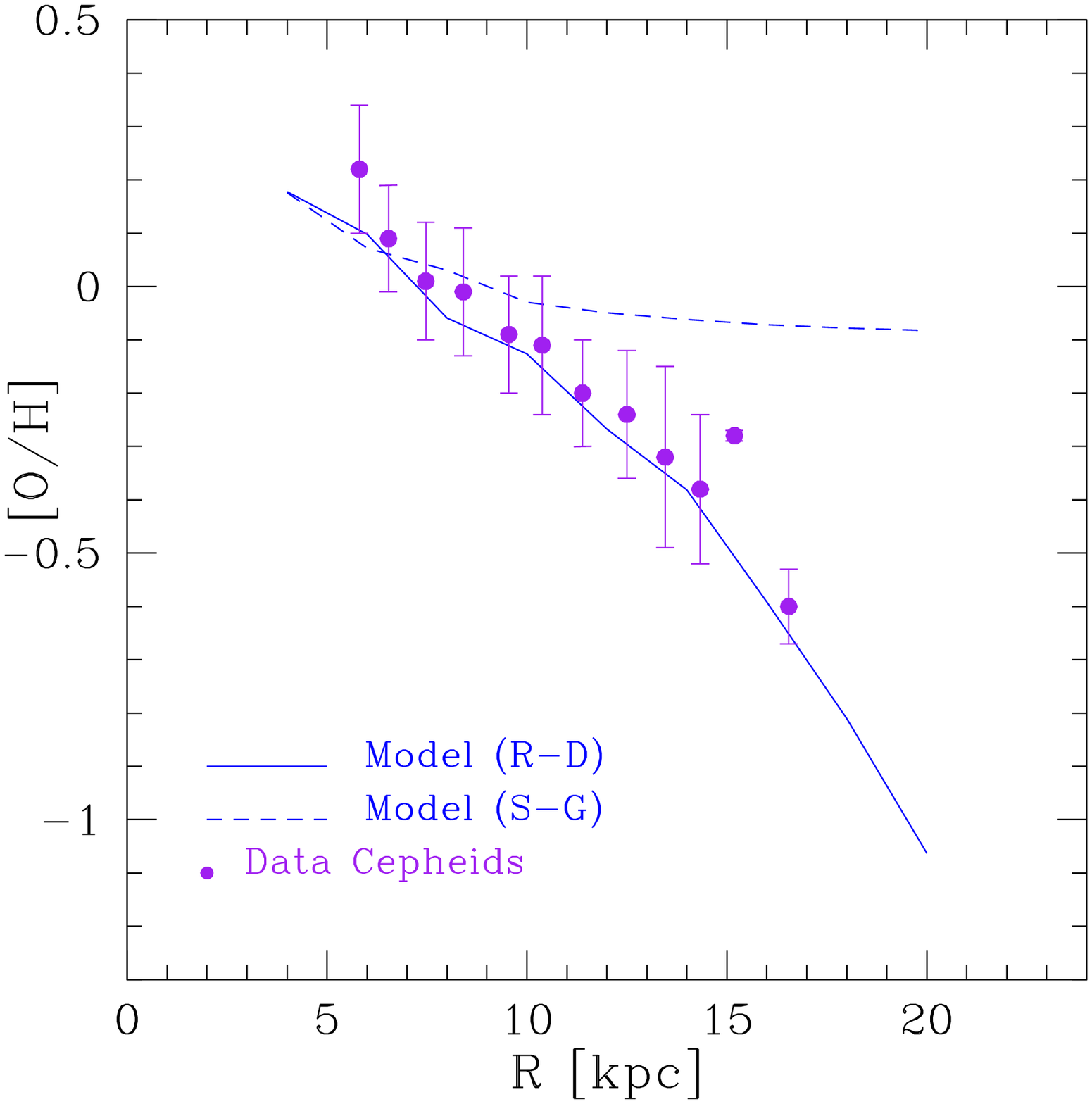} 
    \caption{Oxygen abundance gradient: comparison between
      observations and model (S-G) and (R-D) predictions. The best fit
      in the model (R-D) is obtained using the velocity pattern (III)
      shown in Fig. \ref{patternmott}. The data are taken from Luck \&
      Lambert (2011).}
		\label{NEX_O}
\end{figure}

\begin{figure}
	  \centering   
    \includegraphics[scale=0.4]{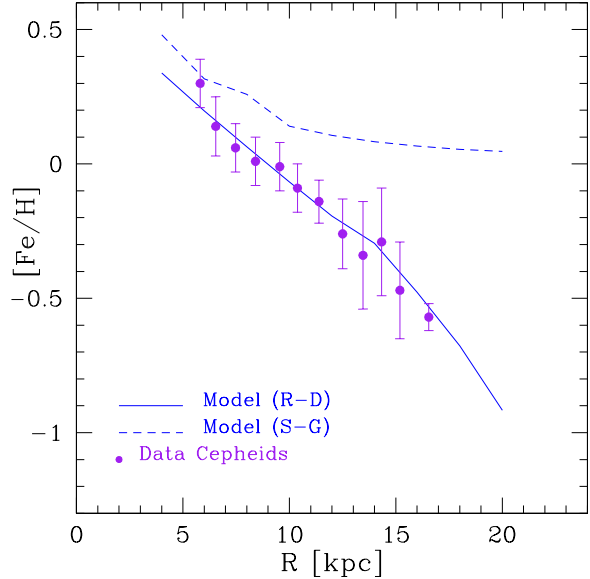} 
    \caption{Iron abundance gradient: comparison between observations
      and model (S-G) and (R-D) predictions. The best fit in the model
      (R-D) is obtained using the velocity pattern (III) shown in
      Fig. \ref{patternmott}. The data are taken from Luck \& Lambert
      (2011).}
		\label{NEX_Fe}
\end{figure} 

In the past, several authors have suggested that the main effect of
taking into account radial gas flows in chemical evolution models is
the steepening of the abundance gradients.  It is interesting to note
that in the majority of previous works, only the O gradient was
discussed in relation to radial flows.  However, in Spitoni \&
Matteucci (2011) it was shown that for a model with just one infall,
aimed at studying just the disk and not the whole Galaxy as the
two-infall model, a linear law for the gas flow velocity could
reproduce both the O and Fe gradient at the same time.  On the other
hand, also in Bilitewski \& Sch\"onrich (2012), who included radial
gas flows in their model, a flat gradient in the central disk region
was found for the iron in some of their models.  The explanation of
this behavior can be found in the combination of several processes:
\begin{itemize}
\item different timescales of production of chemical elements (oxygen
  is produced on short time scales whereas iron on long ones);
\item the two-infall assumption for the gas accretion, in particular
  the assumed total surface mass density distribution for the halo;
\item the inflow velocity pattern considered.
\end{itemize}
In particular, it seems that the flattening of the Fe gradient in the
central disk regions could be due to a combination of variable velocity
for the flow and long timescales for the Fe production from Type Ia
supernovae.
To test our hypothesis, we computed the gradients of
other elements, as shown in Fig. \ref{not_l}, where we report the
predictions for the gradients of O, Fe, C, Ne, and S, obtained with
model R-A. As expected, the elements produced on short time scales,
such as O and Ne show a steeper gradient relative to the static case,
whereas the ones produced on long time-scales, such as C, show the
same behavior as Fe. This means that with that particular velocity
pattern for radial flows (pattern I), the elements produced on long
timescales flatten in the central part of the disk.

\begin{figure}
	  \centering   
    \includegraphics[scale=0.4]{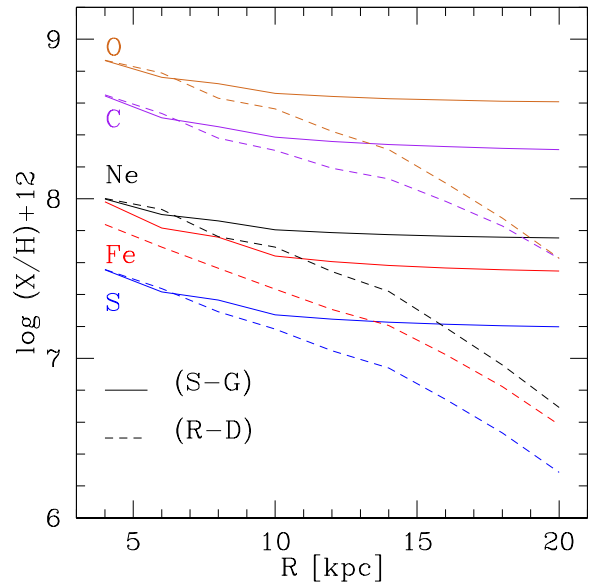} 
    \caption{Present day abundance gradients for O, C, Ne, Fe, and S
      for the static model S-G (solid lines) and for the model with
      radial inflow of gas R-D (dashed lines).}
		\label{SGRD}
\end{figure}

\begin{figure}
	  \centering   
    \includegraphics[scale=0.4]{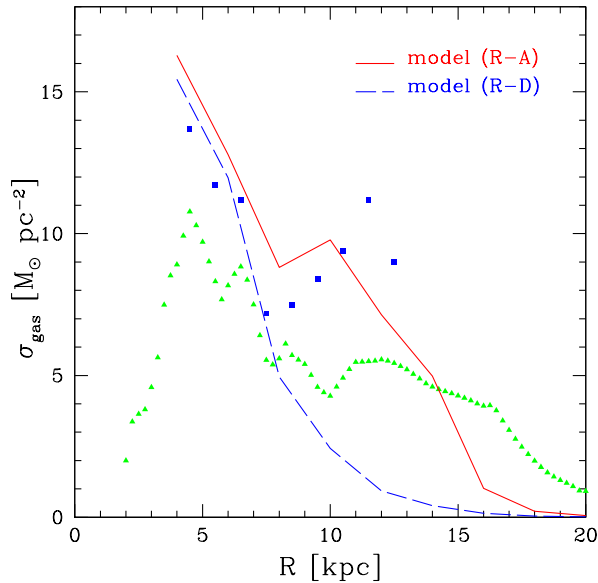} 
    \caption{The surface gas density profile as a function of the
      radial distance for the model (R-D) drawn with the blue dashed
      line and for the model (R-A) drawn with solid red line. The
      green triangles represent the data of Dame et al. (1993),
      whereas the blue squares the data of Rana (1991).}
		\label{GAS_RD_RDA}
\end{figure}

Having said that,
the problem remains that model R-A does not fit the present day Fe
gradient.  This is clearly a peculiarity of the two-infall model since
Spitoni \& Matteucci (2011) could fit both Fe and O gradients at the
same time with a simple linear velocity pattern in the framework of the
one-infall model.  We have then tried another velocity pattern that can fit
the Fe gradient and we have found that, in order to obtain a good fit
to the present day Fe gradient we need to assume for Fe a different
velocity pattern, as shown in Fig. 4 (velocity
pattern II). In particular, we need a strong decrease in the inflow
velocity towards the Galactic centre. With this velocity pattern, the
iron abundance gradient can be very well reproduced, as shown in
Fig. 9. The value of this gradient is:

\begin{equation}
\frac{d\left[\mbox {Fe/H}\right]}{dr}=-0.064\pm  0.004\ \mbox {dex/kpc\ \ (5-17\ kpc)}.
\end{equation}

However, the gradient of O in this case is no more in good agreement
with the data. 

We have then finally tested a model with a total surface mass density
in the halo becoming very small for distances $>$ 10 kpc (model R-D),
a more realistic situation than that in model R-A, and yet a different
velocity pattern (III). In this case we were able to find a very good
agreement for both O and Fe present day gradient, as shown in Figs. 10
and 11. In these figures are shown also the predictions by the
equivalent static model S-G. It is clear that this latter cannot
reproduce the present time gradients because it does not assume any
threshold in the gas density nor radial flows.  In Fig. 12 we report
the present day abundance gradients for O, C, Ne, Fe, and S for the
model R-D with radial flow and for the equivalent static model S-G.
In contrast to model R-A (Fig. 8), we see that in the model R-D all
elements show a steeper gradient compared to the relative static model
S-G, without any flattening in the central parts for elements produced
on long time-scales.  The peculiarity of models R-D and S-G is that
the halo surface mass density is assumed to drop dramatically for
galactocentric distances larger than 10 kpc.  Therefore, the halo
surface mass density distribution seems to play an important role in
shaping the gradients especially in the presence of radial gas flows.
The halo surface mass density can influence, in fact, the gas
distribution in the disk at large galactocentric distances, as already
discussed in Chiappini et al. (2001). Assuming a constant surface
density of 17 M$_{\odot}$ pc$^{-2}$ for the halo is also not realistic
for distances larger than 10 kpc where the surface mass density of the
disk is much smaller. So, our best model is model R-D.

\begin{figure}
	  \centering   
    \includegraphics[scale=0.4]{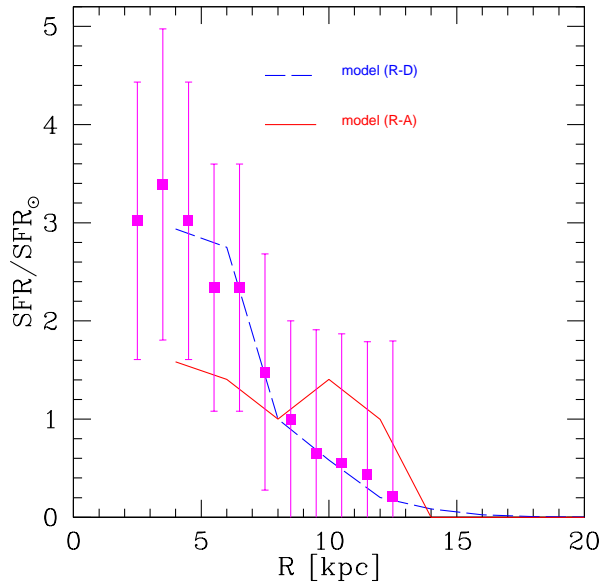} 
    \caption{ The SFR normalized to the solar values
      as a function of the galactocentric distance for the (R-A) and
      (R-D) models.  The data are taken by Rana (1991) and are
      reported with the magenta points and relative error bars.}
		\label{SFR_RANA}
\end{figure}

\begin{figure}
	  \centering   
    \includegraphics[scale=0.4]{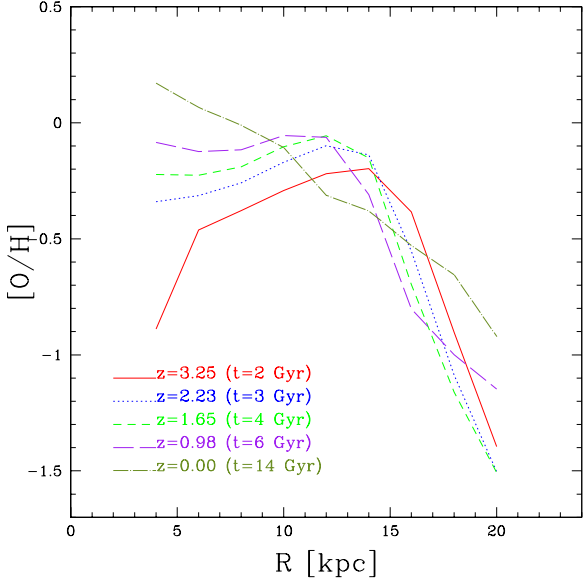} 
    \caption{Evolution with redshift of the abundance gradients for
      the model R-A for the Oxygen.  Model R-A assumes a threshold,
      Inside-Out formation, constant SFE and radial flows.  The
      evolution is studied by computing the abundance gradients for
      the redshifts z=3.25, 2.23, 1.65, 0.98 and 0.0 whose correspond
      to the times t=2, 3, 4, 6 and 14 Gyr after the Big Bang (in a
      $\Lambda CDM$ cosmology).}
		\label{altoredshiftORA}
\end{figure}

 We pass now to analyze how the profile of the gas along the
  Galactic disk is modified with the velocity patterns used for the
  best model (R-D) and the model (R-A). In Fig. 13, we compare our
  results for the gas surface density profile as a function of the
  radial distance with the data of Dame et al. (1993) and Rana (1991).
  As stated in Spitoni \& Matteucci (2011), the main effect of the
  radial inflow with a variable velocity is to concentrate the gas in
  the inner regions of the Galaxy. In the inner regions between 5 and
  8 kpc our results are in very good agreement with the data of Rana
  (1991).  On the other hand, both the models (R-D) and (R-A)
  underestimate the gas density in the outer regions, but the model
  (R-A) shows agreement with the data of Rana (1991) between 8 and 14
  kpc.  Our models, like the majority of the others, are not able to
  reproduce the decrease in gas density for R $\leq$ 5 kpc of Dame et
  al. (1993). This is probably because the effect of bar is not
  considered.

 In Fig. 14 the SFR along the disk of the Galaxy for the (R-D) and
  (R-A) models is compared with the data of Rana (1991).  In this
  figure we show the SFR normalized to the solar value as a function
  of the Galactocentric distance. Because of the uncertainties in this
  data set, as it can be seen from the large error bars, we cannot
  draw firm conclusions, although the model (R-D) seems to fit the
  data very well. On the other hand,  model (R-A) underestimates the
  $SFR/SFR_{\odot}$ for R $\leq$ 6 kpc, but  its results are consistent with observations
  within the error bars given by one standard deviation.

\subsection{The high redshift gradients}

We show in Figs. 15 and 16 the temporal evolution of the gradients for
the model (R-A) and the model with variable SFE (R-B) for the oxygen.
Concerning the R-A model (that assumes a constant SFE) at z=3,
corresponding to a cosmic time of 2 Gyr in a $\Lambda$CDM cosmology,
the gradient reveals the inversion of the gradient, i.e. a positive
value in the inner regions of the disk. Then the gradient exhibits a
steepening in time, mostly in the inner regions, and at z=0.98 the
gradient inversion disappears, giving way to a plateau in the inner
parts of the disk. After that, the steepening continues until the
gradient reaches the known shape at the present time.  If we look at
the temporal evolution for the model R-B, it is evident that by
assuming a variable SFE, the gradient inversion is never
predicted. However, as in the case of the R-A model, the steepening
with time is still present, and this is due to the threshold in the
gas density.

\begin{figure}
	  \centering   
    \includegraphics[scale=0.4]{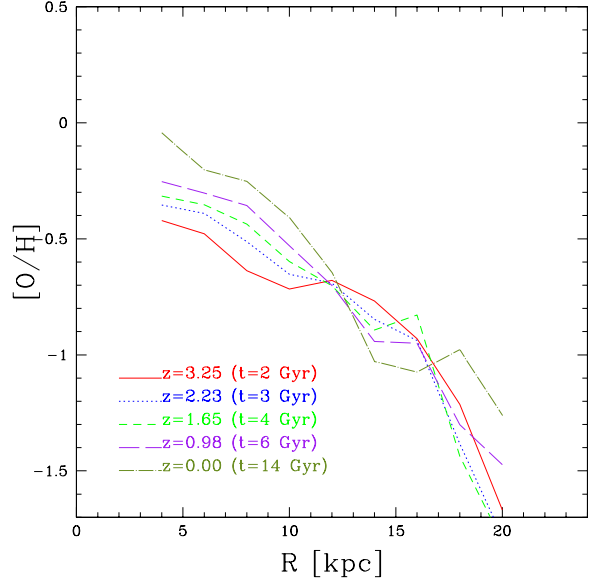} 
    \caption{Evolution with redshift of the abundance gradients for
      the model R-B for the Oxygen.  Model R-B is the same of the one
      R-A but with variable SFE.  The evolution is studied by
      computing the abundance gradients for the redshifts z=3.25,
      2.23, 1.65, 0.98 and 0.0 whose correspond to the times t=2, 3,
      4, 6 and 14 Gyr after the Big Bang (in a $\Lambda CDM$
      cosmology).}
		\label{altoredshiftORB}
\end{figure}

Finally, we present in Fig. 17 the temporal evolution for our best
model R-D. We recall that this model is able to fit the present day
abundance gradients both for O and Fe by means of the same pattern of
gas flow velocity (pattern III in Figure 4).  Moreover, at z=3 the
model presents the inversion of the gradient, but less pronounced than
the model R-A. As expected, the abundance gradient steepens in time.
Using the same models listed in Table 3, we derived the abundance
gradients at the time t = 2 Gyr. The results are shown in Table 4
where are reported the theoretical abundance gradients predicted by
our models and their uncertainties. They were computed by means of the
least square method, as done for the present day gradients.  In the
fitting procedure, we considered only the inner range of the thin disk
(2-8 kpc) for a better comparison with the data by Cresci et
al. (2010).

\begin{figure} 
	  \centering   
    \includegraphics[scale=0.4]{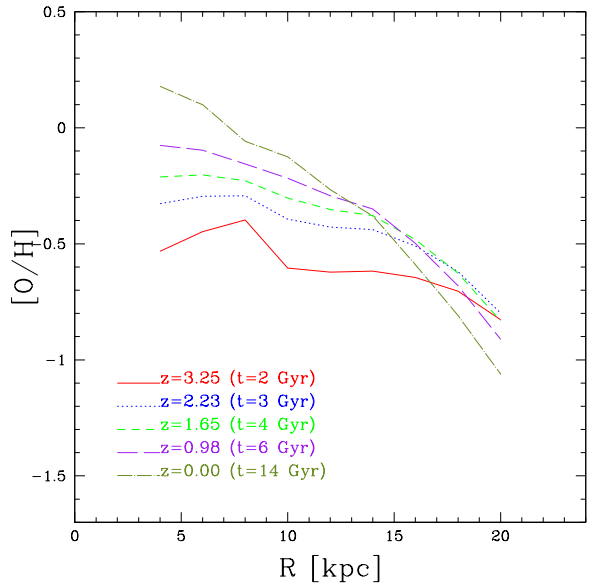} 
    \caption{Evolution with redshift of the abundance gradients for
      the best model R-D for the oxygen.  The evolution is studied by
      computing the abundance gradients for the redshifts z=3.25,
      2.23, 1.65, 0.98 and 0.0 whose correspond to the times t=2, 3,
      4, 6 and 14 Gyr after the Big Bang (in a $\Lambda CDM$
      cosmology).}
		\label{altoredshiftORD}
\end{figure}

\begin{table} 
\caption{Abundance gradients at high redshift ($z=3$) predicted by
  models S-A, S-B, S-C, S-E, S-G, R-A, R-B, R-C, and R-D. The
  gradients were computed with the least square method and only in the
  inner parts of the thin disk, where the trend is approximated by a
  linear regression.}
\label{ta4}
\begin{center}
\begin{tabular}{cccc}
  \hline\hline
\noalign{\smallskip}

 Model& Range  & $\frac{d\left(log(O/H)+12\right)}{dRg}$&$\frac{d\left(log(Fe/H)+12\right)}{dRg}$ \\

 & [kpc]  &[dex kpc$^{-1}$]&[dex kpc$^{-1}$] \\
\noalign{\smallskip}
  \hline 
  SA & 2-8 & $+0.036\pm0.009$&$+0.047\pm0.008$\\
 SB & 4-8  & $-0.066\pm0.008$&$-0.066\pm0.003$\\
 SC & 2-8 &  $+0.038\pm0.007$&$+0.045\pm0.009$\\
 SE & 4-8 &  $+0.022\pm0.004$&$+0.034\pm0.001$\\

 SG & 4-8 &  $+0.039\pm0.004$&$+0.060\pm0.001$\\

RA & 4-8 & $+0.128\pm0.050$&$+0.060\pm0.002$\\

RB & 4-8 & $-0.054\pm0.014$&$-0.053\pm0.014$\\

RC & 4-8 & $+0.035\pm0.002$&$+0.049\pm0.003$\\
RD & 4-8 &  $+0.034\pm0.049$&$+0.050\pm0.005$\\

  \hline
 \end{tabular}
\end{center}
\end{table}

As it is evident from Table 4, if we look at the abundance gradients
at high redshift, both iron and oxygen exhibit a gradient inversion in
the inner part of the disk. In other words, if we consider for example
the gradients of our best model (R-D, red solid lines), at t = 2 Gyr,
they show an increase of metallicity from the outer regions up to 10
kpc where they reach a peak and then a decrease for R $<$ 10 kpc.
This is not true for the model (S-B) with variable SFE: this model
predicts a negative abundance gradient even at high redshift.  This
theoretical gradient inversion in the innermost part of the disk is
confirmed by the data taken by Cresci et al. (2010) (only for the
oxygen because only for this element there are available metallicity
measurements), which found positive gradients for both of the galaxies
considered (see Table 2).  Our high redshift positive gradients are in
general smaller than the observed ones: the reason for this could
reside in the nature of the sources used to derive the metallicity at
high redshift. In fact, in the sample by Cresci et al. (2010), there
are Lyman-Break galaxies  showing evidence for disk.  However
these galaxies might be large objects than the Milky Way and this
could be the reason for our gradients being smaller.

Our explanation for the gradient inversion in the Milky Way is based
on the inside-out disk formation:

\begin{itemize}
\item at early epoch (t=2 Gyr from the Big Bang, z = 3) the efficiency
  of chemical enrichment (i.e. of the SFR), in the
  inner regions is high but the rate of infalling primordial gas is
  dominating, thus diluting the gas more in the inner regions than in
  the outer ones;
\item as time passes by, the infall of pristine gas in the inner parts
  decreases and the chemical enrichment takes over;
\item then, at later epochs, the SFR in the inner
  regions is still much higher than in the outer parts of the disk
  where the gas density is very low, but the infall is lower and the
  abundance gradients assume the shape seen in the observational data.

\end{itemize}

 Since we assumed a flow velocity constant in time we checked the
  robustness of our results for the (R-D) model concerning the
  inversion of gradient at redshift z=3, considering a time-dependent
  velocity pattern consistent with the present time pattern III  introduced in
  Fig. 4, and with the work of Bilitewski
  \& Sch\"onrich (2012).

 We assume for this test the extreme case where the pattern III is the
 present day law, and we look for the pattern at times $t \leq 2$ Gyr
 which could end up as pattern III.  In Fig. 3 of Bilitewski \& Sch\"onrich
 (2012), time dependent velocity patterns at times $t=$ 3 Gyr, 6 Gyr, 9
 Gyr, and 12 Gyr are presented. We find that the radial velocity pattern
 at 3 Gyr is related to the one at 12 Gyr by the following expression:
 $v_r(3 \ Gyr)=1.37 \times v_r(12 \ Gyr)$.

Therefore, for the model (R-D) we assume that the pattern at the early times
up to 2 Gyr  follows this
equation:
\begin{equation}
v_r(t \leq 2
\ Gyr)=1.5 \times v_r (pattern III).
\end{equation}
 In Fig. 18, we labeled this velocity profile as ``pattern V''. In
 Fig. 19 we show the abundance gradient for the oxygen at t=2 Gyr
 using this new velocity pattern, and we can see that the inversion of
 the gradient is preserved. We consider this a robust test.

 On the other hand, a full  treatment of a time dependent velocity pattern is not
 the aim of this paper and it will be discussed in a future work.

The last test was to check a smaller velocity of the flow at 2 Gyr (pattern IV in
Fig. 18). In Fig. 19 we can see that also in this case  a gradient
inversion is present.

\begin{figure}
	  \centering   
    \includegraphics[scale=0.4]{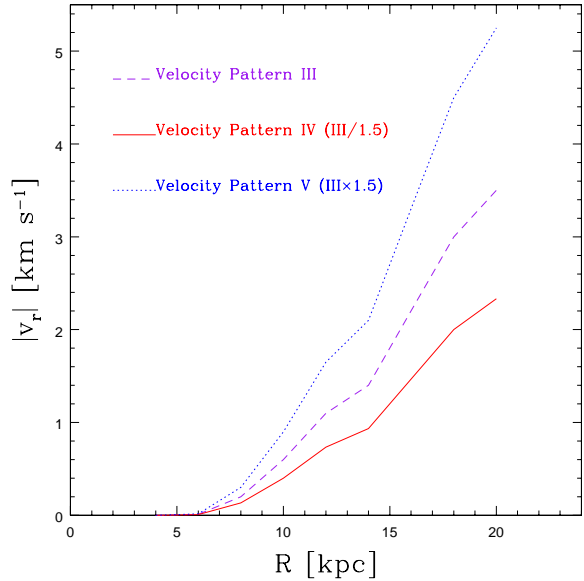} 
    \caption{Radial inflow velocity used to check the inversion of
      gradient at high redshift: velocity pattern III (dashed magenta
      line), IV (red solid line), V (blue dotted line). We applied
      these patterns to the model (R-D) up 2 Gyr. }
		\label{}
\end{figure}

\begin{figure}
	  \centering   
    \includegraphics[scale=0.4]{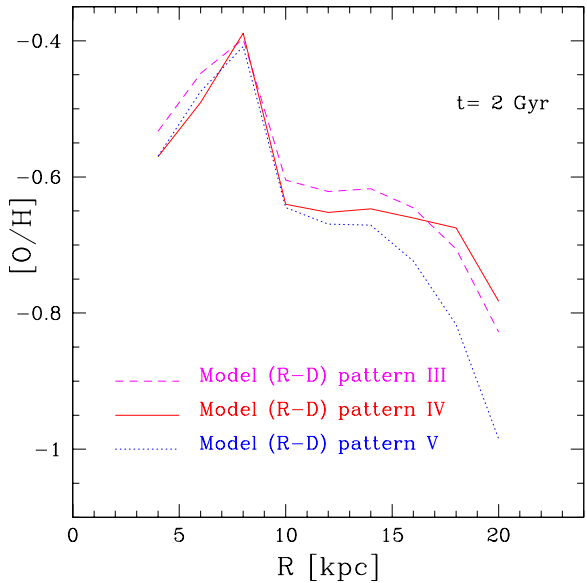} 
    \caption{(R-D) model predictions for Oxygen abundance gradient at
      2 Gyr using the velocity pattern III (dashed magenta line), IV
      (red solid line), V (blue dotted line) shown in Fig. 18.}
		\label{}
\end{figure}

\section{Conclusions}
We have computed the chemical evolution of the Milky Way with
particular attention to the formation and evolution of the abundance
gradients. We have tested the effects of several physical processes
which can influence the formation of gradients, and studied the
temporal evolution of gradients.  In particular, we have tested
  whether the gradient inversion of redshift z=3 is preserved in the
  presence of radial gas flows, and this is the main novelty of the
  paper.

 Our main conclusions can be summarized as follows:

\begin{itemize}
 \item If we do not assume an inside-out formation for
the thin disk ($\tau_D$ constant), the present day abundance gradients
provided by the model are too flat in the inner part of the disk
and the observations are not reproduced (model S-E).

\item Without a threshold in the standard static model
by Chiappini et al.(2001) even if an inside-out formation is 
considered (model S-C), the gradient is too flat
between 6-12 kpc and, mostly in the outer parts of the disk, the
observational data are not reproduced. The threshold steepens
the gradient in the outermost regions.   

\item To reproduce the abundance gradients in the whole disk, it is
  not possible to assume a variable SFE (model
  S-B) because, with this assumption, a flat gradient in the outer
  part of the thin disk is necessarily obtained (R $>$ 14 kpc). The
  use of a constant $\nu$ = 1 Gyr$^{−1}$ along the whole disk is the
  best choice to reproduce the chemical evolution of the Milky Way.

\item 
The best model we found (R-D) assumes inside-out formation of the
disk, a constant efficiency of SF along the disk, a surface halo mass
density decreasing with galactocentric distance and radial gas flows
with a velocity pattern predicting a decrease of the speed of the gas
flow for decreasing galactocentric distance in the range between 0 and
3.8 km s$^{-1}$.

\item We found that the assumed velocity pattern for radial gas flows
  is a crucial parameter in shaping the chemical evolution of the disk
  and in particular the abundance gradients and their evolution.

\item All the chemical evolution models presented here show a steepening
of the gradient in time, mostly in the inner regions.

\item At z=3 a gradient inversion in the inner regions of the
  thin disk is predicted for most of our chemical evolution models,
  including the best one.   With a time dependent velocity
    radial inflow, the inversion of the abundance
    gradient at early times is preserved.

This inversion is observationally confirmed (Cresci et
  al. 2010) and is interpreted as due to the strong infall of
  primordial material at early times in the innermost disk regions, as
  predicted by the inside-out formation. Therefore, the gradient
  inversion is a direct consequence of the inside-out disk formation.

\item The only model with inside-out that does not predict the
  gradient inversion at high redshift is that with variable SFE (model
  S-B). In this model the gradient at z = 3 is negative as the present
  day one. If the gradient inversion will be confirmed, we can
  conclude that a variable the SFE, which
  contrasts the effect of the infall at early times, is unlikely.

\item Finally, a word of caution is necessary relative to the observed
  gradient inversion at high z; Yuan et al. (2003) showed that the
  measured metallicity gradient changes systematically with angular
  resolution and annular binning. Seeing-limited observations produce
  significantly flatter gradients than higher angular resolution
  observations. For these reasons, more observations are needed in the
  future to confirm Cresci et al. (2010) high redshift positive inner
  gradients.

\end{itemize}

 It is worth, before concluding, to discuss how robust are the
  above conclusions.  We run several models changing the main
  assumptions which are: inside-out formation of the disk, efficiency
  of star formation, radial flows and their speed, threshold gas
  density for star formation and distribution of the total surface
  mass density of the Galactic halo. The inside-out formation scenario
  is a well tested assumption suggested by the majority of papers
  concerning the Galactic disk (e.g. Pilkington et al. 2012) and also
  by observations at high redshift (e. g. Mateo-Munoz et al. 2007). 
We  conclude that the inside-out formation is necessary to reproduce the
  abundance gradients but also to obtain a gradient inversion at high
  redshift and this is a robust conclusion. In fact, even changing all
  the other physical processes, with the exception of a variable star
  formation efficiency with galactocentric distance, a model with
  inside-out always predict such an inversion. 

On the other hand, other assumptions are probably still debatable,
such as the halo total surface mass density distribution. This might
appear like a quantity which should not influence abundance gradients
in the disk. Instead, this quantity has a noticeable influence on the
disk gradients at large galactocentric distances. In fact, at such
distances the total surface mass density of the disk is negligible and
that of the halo might predominates. This fact was already discussed
in Chiappini et al. (2001). We find that both the O and Fe gradient at
the present time in the disk can be well reproduced if the halo total
surface mass density distribution is a strong decreasing function of
the galactocentric radius, but we cannot say if this is an unique
solution.

  A variable efficiency of
  star formation with radius can be certainly ruled out if the
  gradient inversion will be confirmed and this is also a robust
  result. 

On the other hand, the presence or absence of a gas density
  threshold cannot be firmly established on the basis of our models,
  although in the presence of radial flows we obtain better results
  without a threshold.  

Finally, the model could be further improved
  by considering stellar migration in the disk which will be the
  subject of a forthcoming paper.

\section*{Acknowledgments}
E. Spitoni and F. Matteucci aknowledge financial support from PRIN
MIUR 2010-2011, project “The Chemical and dynamical Evolution of the
Milky Way and Local Group Galaxies”, prot. 2010LY5N2T. We thank the
anonymous referee for his/her suggestions which improved the paper.

\end{document}